\UseRawInputEncoding
\documentclass[aip, apl, amsmath,amssymb, reprint,]{revtex4-2} 

\usepackage{graphicx}
\usepackage{dcolumn}
\usepackage{bm}

\usepackage[utf8]{inputenc}
\usepackage[T1]{fontenc}
\usepackage{mathptmx}
\usepackage{etoolbox}
\usepackage{siunitx}
\usepackage[usenames,dvipsnames]{color}

\makeatletter
\def\@email#1#2{%
 \endgroup
 \patchcmd{\titleblock@produce}
  {\frontmatter@RRAPformat}
  {\frontmatter@RRAPformat{\produce@RRAP{*#1\href{mailto:#2}{#2}}}\frontmatter@RRAPformat}
  {}{}
}%
\makeatother

\begin{document}

\title{Experimental evidence of dominant ultrafast diffusive energy transport by hot electrons in Cu}

\author{J.~Jarecki}
\affiliation{Max-Born-Institut f\"ur Nichtlineare Optik und Kurzzeitspektroskopie, 12489 Berlin, Germany}
\email{jarecki@mbi-berlin.de}

\author{L.~Mehner}
\affiliation{Institut f\"ur Physik und Astronomie, Universit\"at Potsdam, 14476 Potsdam, Germany}

\author{M.~Mattern}
\affiliation{Max-Born-Institut f\"ur Nichtlineare Optik und Kurzzeitspektroskopie, 12489 Berlin, Germany}
\email{mattern@mbi-berlin.de}

\author{A.~Jurgilaitis}
\affiliation{MAX IV Laboratory, Lund University, PO Box 118, 221 00 Lund, Sweden}

\author{S. P.~Zeuschner}
\affiliation{Institut f\"ur Physik und Astronomie, Universit\"at Potsdam, 14476 Potsdam, Germany}

\author{B.~Ahn}
\affiliation{MAX IV Laboratory, Lund University, PO Box 118, 221 00 Lund, Sweden}

\author{F.~Baltrusch}
\affiliation{Institut f\"ur Physik und Astronomie, Universit\"at Potsdam, 14476 Potsdam, Germany}

\author{J. C.~Ekstr\"om}
\affiliation{MAX IV Laboratory, Lund University, PO Box 118, 221 00 Lund, Sweden}

\author{D.~Kroon}
\affiliation{MAX IV Laboratory, Lund University, PO Box 118, 221 00 Lund, Sweden}

\author{M.~Herzog}
\affiliation{Institut f\"ur Physik und Astronomie, Universit\"at Potsdam, 14476 Potsdam, Germany}

\author{C.~Walz}
\affiliation{Institut f\"ur Physik und Astronomie, Universit\"at Potsdam, 14476 Potsdam, Germany}

\author{F.-C.~Weber}
\affiliation{Institut f\"ur Physik und Astronomie, Universit\"at Potsdam, 14476 Potsdam, Germany}
\affiliation{Helmholtz-Zentrum Berlin f\"ur Materialien und Energie GmbH, Wilhelm-Conrad-R\"ontgen Campus, BESSY II, 12489 Berlin, Germany}

\author{J.~Larsson}
\affiliation{Department of Physics, Lund University, PO Box 118, Lund 221 00, Sweden}
\affiliation{MAX IV Laboratory, Lund University, PO Box 118, 221 00 Lund, Sweden}
\affiliation{LINXS Institute of advanced Neutron and X-ray Science, Mesongatan 4, 224 84 Lund, Sweden}

\author{M.~Hehn}
\affiliation{Institut Jean Lamour (UMR CNRS 7198), Universit\'e Lorraine,  54000 Nancy,   France}

\author{J.-E.~Pudell}
\affiliation{European X-ray Free-Electron Laser Facility, 22869 Schenefeld, Germany}

\author{D.~Schick}
\affiliation{Max-Born-Institut f\"ur Nichtlineare Optik und Kurzzeitspektroskopie, 12489 Berlin, Germany}

\author{A.~von Reppert}
\affiliation{Institut f\"ur Physik und Astronomie, Universit\"at Potsdam, 14476 Potsdam, Germany}

\author{M.~Bargheer}
\affiliation{Institut f\"ur Physik und Astronomie, Universit\"at Potsdam, 14476 Potsdam, Germany}
\affiliation{Helmholtz-Zentrum Berlin f\"ur Materialien und Energie GmbH, Wilhelm-Conrad-R\"ontgen Campus, BESSY II, 12489 Berlin, Germany}
\date{\today} 

\begin{abstract}
When the dimensions of structures shrink to the order of the inelastic mean free path of the energy-carrying quasi-particles, the character of energy transport changes from diffusive to ballistic.
However, the point of transition remains a matter of debate.
Here, we determine the dominant channel of energy transport through a nanoscale Cu layer as a function of its thickness.
The energy rapidly transferred across Cu via hot electrons from a photo-excited Pt layer into a buried Ni detection layer translates into a rapid expansion of the Ni layer probed via ultrafast x-ray diffraction.
The non-linear dependence of the Ni strain amplitude on the absorbed laser fluence indicates that the transport through Cu becomes more efficient with increasing fluence.
This fluence-dependent transport efficiency is reproduced by a diffusive energy transport model and serves as a generally applicable experimental approach to distinguish diffusion from ballistic transport.
Following this approach, we identify diffusive electronic energy transport to govern the spatial energy distribution for Cu layer thicknesses larger than twice the electronic inelastic mean free path.
\end{abstract}

\maketitle
The response of solid state materials to ultrafast optical excitation is determined by the initial energy distribution and the subsequent energy dissipation on femto- to nanosecond-timescales.
This makes understanding energy transfer processes essential not only from a fundamental point of view, but also vital for transferring solid state physics like ultrafast magnetism~\cite{koop2010,stei2024,ishibashi2025,xu2017,hennecke2025,mattern2024,ahn2022,rott2006} and plasmon-driven chemistry~\cite{sarhan2019,stete2025,verm2024} to every day applications.

A femtosecond laser pulse introduces non-thermal occupations of quantum states and non-equilibria among different degrees of freedom~\cite{carpene2006,waldecker2016}, modifying the local energy transfer among these different subsystems~\cite{carpene2006,wilson2020}. 
This substantially modifies the local energy transfer among the different subsystems~\cite{carpene2006,wilson2020} as well as the energy transport probed on nanometer length and picosecond timescales compared to thermal equilibrium~\cite{romeo2025,wang2012,roth2019,heck2023}.
In purely metallic systems, the photo-excited electrons --- initially out of equilibrium with the other degrees of freedom --- govern the spatial distribution of heat~\cite{bloc2019,wang2012} until their energy is transferred to the remaining subsystems typically dominated by the phonons due to their large heat capacity.
Sophisticated spatial energy distributions can be tailored by combining nano-scaled metal layers with different electron-phonon coupling strengths and electronic heat capacities~\cite{stete2025,pude2018,choi2014}.
This opens new pathways to control the spatial energy profile after electron-phonon equilibration by heterostructure design~\cite{wang2024,pude2020,matt2022,igar2023,shin2020} beyond tuning the optical absorption profile~\cite{seibel2022}.
Consequently, metallic heterostructures are a particularly versatile material class enabling control over energy transfer processes and related ultrafast dynamics.

If the dimensions of the metal layers shrink towards the order of the energy-dependent inelastic mean free path of the electrons, the nature of the electronic energy transport is expected to change from diffusive to ballistic~\cite{chen2001,heck2023}.
The point of transition, however, remains a topic of discussion and ongoing debate.
Following the pioneering work of Brorson et al.~\cite{bror1987}, several experiments claimed that ballistic or super-diffusive electrons dominate the energy transport through Cu or Au of thicknesses up to \qty{300}{nm} -- far beyond the electronic inelastic mean free paths -- based on the linear relation between the delay of the transient response of a buried layer normalized to its maximum and the thickness of the noble metal propagation layer~\cite{berg2016,berg2020,fert2019}.
However, recent studies demonstrated that the observed linear space-time relation is captured by completely diffusive electronic energy transport~\cite{jang2020,pude2020,wils2017}.
Although direct experimental evidence of ballistic and superdiffusive electrons was provided~\cite{heck2023,jechumtal2024,seifert2023,rouzegar2025} it remains unclear to what extent they significantly contribute to ultrafast energy transfer after photo-excitation~\cite{ashok2025,jechumtal2024,sivan2024,hohlfeld2000}.

In this letter, we investigate electronic energy transport in Cu sandwiched between a Pt absorption layer and a Ni detection layer by monitoring the transient strain of the buried Ni layer via ultrafast hard-x-ray diffraction (UXRD) for varying Cu thicknesses.
Instead of on the timescale of the electronic energy transport, our approach is based on analyzing the ratio between the prompt Ni expansion immediately after photoexcitation and the delayed expansion after thermal equilibration, that is a measure of the fraction of the optically deposited energy that is rapidly transported across the Cu layer into Ni.
We observe that the relative amount of rapidly transferred energy increases with increasing fluence which becomes more pronounced with increasing Cu layer thickness.
This observation is reproduced by a purely diffusive two-temperature model (d2TM) and not by a model that includes ballistic electronic transport.
Consequently, the fluence dependence of the electronic energy-transport efficiency -- manifested as a nonlinear increase in the rapidly transmitted energy -- provides a generally applicable means to distinguish diffusive from ballistic electronic transport.
Based on this analysis, we find that hot-electron diffusion governs energy transfer across Cu layers thicker than \qty{50}{nm}, i.e. roughly twice the electronic inelastic mean free path.

\begin{figure}[!ht] 
    \centering
    \includegraphics[width=1\columnwidth]{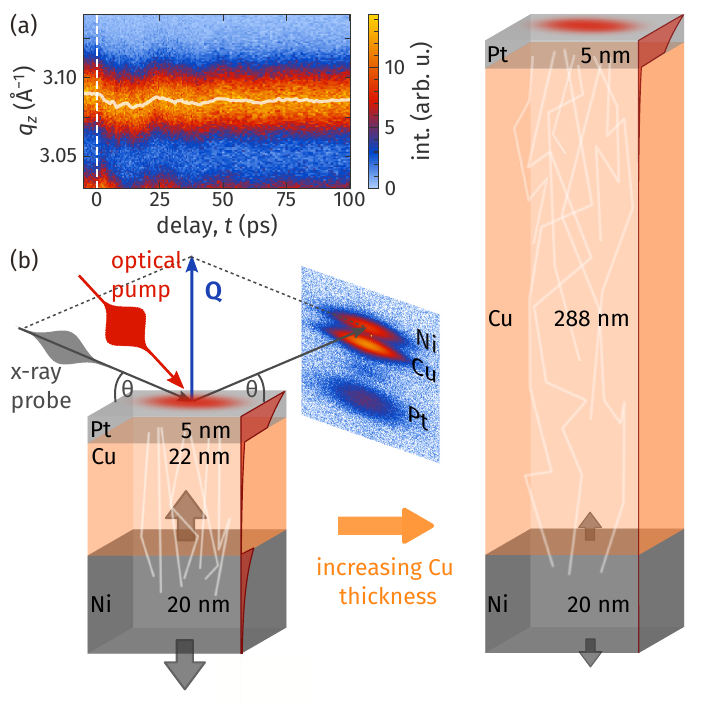}
    \caption{Schematic of the UXRD experiment on the Pt-Cu-Ni heterostructures.
    The metallic heterostructures are excited by $800\,\text{nm}$ femtosecond laser pulses and we monitor the shift of the (111) Bragg peak of the Ni detection layer using hard-x-ray pulses incident under the Bragg angle of Ni ($22.1^\circ$) on a pixelated position-sensitive area detector. 
    Transforming the diffraction pattern on the detector into reciprocal space yields the diffracted intensity along the out-of-plane reciprocal coordinate $q_z$ as function of the pump-probe delay $t$ exemplified for the sample with $22\,\text{nm}$ Cu (a).
    (b) While for $22\,\text{nm}$ of Cu a significant part of the pump pulse is absorbed by Ni layer (red shaded area), for Cu layers thicker than $70\,\text{nm}$ the direct absorption in Ni is negligible and Ni is only indirectly excited via electronic energy transport (see Fig.~S5).}
    \label{fig:setup}
\end{figure}

We study the ultrafast energy transport via electrons through Cu embedded within a metallic heterostructure.
The Cu layer thicknesses ranging from the inelastic mean free path ($\approx25\,\text{nm}$)~\cite{schmuttenmaer1994} of the electrons up to an order of magnitude larger covering the whole range of discussion.
The metallic heterostructures built of \qty{5}{nm} Pt capping layer, the Cu transport layer of \qty{22}{nm}, \qty{47}{nm}, \qty{136}{nm}, and \qty{288}{nm} thickness, and a buried \qty{20}{nm} Ni detection layer (see Fig.~\ref{fig:setup}(b)) were sputtered onto a glass substrate with a thin Ta buffer layer to facilitate textured growth with the preferred (111) orientation.
In our time-resolved experiments, the Pt-absorption layer initially localizes the optically deposited energy (see red shaded areas in Fig.~\ref{fig:setup}(b) and Fig.~S5) and serves as a heat source.
Its thickness of only \qty{5}{nm} ensures a negligible effect of transport within the Pt layer~\cite{berg2020}.
Instead, the transport properties of electrons and phonons of the Cu propagation layer determine the rapid spatial energy redistribution within the heterostructure.
We probe this energy transfer by the expansion of the buried Ni detection layer, which is directly proportional to the ultrafast heating~\cite{matt2023}.
In order to characterize the energy transport, we measured each sample at a low and high incident excitation fluence $F$ with a fixed ratio (see supplementary material).

The experiments were mainly conducted at the FemtoMAX beamline at the MAX IV laboratory~\cite{enqu2018}, and complemented by additional UXRD measurements at a table-top laser-driven plasma x-ray source (PXS)~\cite{zamp2009,schi2012}.
There is no conceptual difference between both setups and we verified the comparability of both experiments by calibration measurements (see Fig.~S1).
At both setups, the samples were excited by $<100\,\text{fs}$ (FWHM) laser pulses with a central wavelength of $800\,\text{nm}$.
We studied the laser-induced lattice response of the buried Ni detection layer, by tracking its Bragg peak position on a pixelated position-sensitive \textsc{Pilatus} area detector via reciprocal space slicing~\cite{zeus2021} employing femtosecond hard x-ray pulses with $8\,\text{keV}$ photon energy (see Fig.~\ref{fig:setup}).
The measurements carried out at the Bragg angle of Ni at $22.1^\circ$ also provide access to the dynamics of the Cu Bragg peak (see the resulting strain dynamics in Fig.~S4).

In order to extract the layer-specific picosecond strain response, we first transform the detector images into reciprocal space~\cite{krieg2012} and integrate along the in-plane coordinates $q_y$ and $q_x$, which yields the diffracted intensity along $q_z$ as a function of the pump-probe delay $t$ displayed for the sample with $d_\text{Cu}=22\,\text{nm}$ Cu in Fig.~\ref{fig:setup}(a).
We determine the Ni and Cu Bragg peak positions $Q_z^\text{Ni}(t)$ and $Q_z^\text{Cu}(t)$ (see solid white line in Fig.~\ref{fig:setup}(a)) by fitting the diffracted intensity by a Pseudo-Voigt profile.
Finally, the relative change of the Bragg peak position with respect to its position before excitation $Q_{z,0}^\text{Ni}$ denotes the transient strain $\eta_\text{Ni}(t)=-\left(Q_z^\text{Ni}(t)-Q_{z,0}^\text{Ni}\right)/Q_{z}^\text{Ni}(t)$.

\begin{figure}[!ht]
    \centering
    \includegraphics[width = 8.5cm]{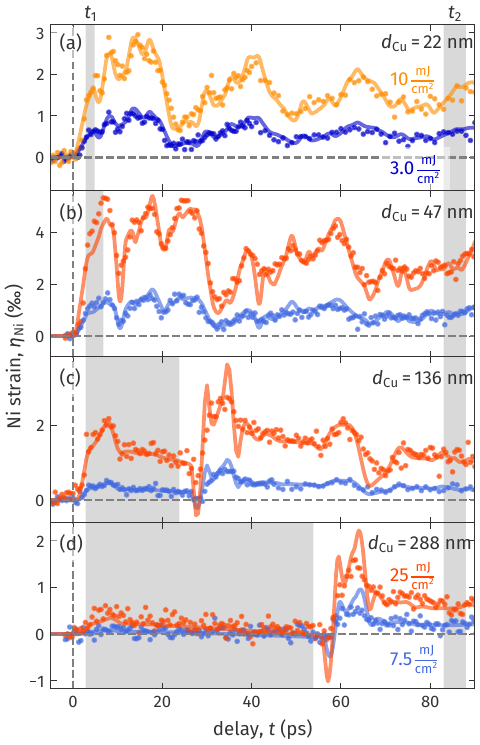}
    \caption{\label{fig:fig_2_delayscans}
    The fluence-dependent strain response of the Ni layer (symbols) extracted from the transient Bragg peak position for the different samples of Cu thicknesses from \qty{22}{nm} (a), \qty{47}{nm} (b), \qty{136}{nm} (c) and \qty{288}{nm} (d).
    Solid lines display the modeling in the framework of a diffusive two-temperature model (d2TM) explicitly considering the effect of the initial electron-phonon nonequilibrium on the electronic heat conductivity.
    All measurements were conducted at \qty{7.5}{mJ\per cm\squared} (blue) and \qty{25.0}{mJ\per cm\squared} (red) except for the additional measurements at the PXS on the \qty{22}{nm} sample, which was measured at \qty{3.0}{mJ\per cm\squared} (darkblue) and \qty{10.0}{mJ\per cm\squared} (orange).
    The gray shaded areas mark the delay ranges used to calculate the fluence response parameter $A$ according to Eq.~\eqref{eq:nonlinearity}.
    }
\end{figure}

We observe a rapid expansion of Ni (positive strain) within the first picoseconds after excitation (symbols in Fig.~\ref{fig:fig_2_delayscans}) irrespective of the Cu layer thickness.
It is followed by coherent oscillations in the strain that originate from the strain pulse launched at the surface by the fast expansion of the photo-excited Pt layer.
This strain pulse enters Ni at delays $t>d_\text{Cu}/v_\text{s,Cu}$ (right border of first gray shaded areas in Fig.~\ref{fig:fig_2_delayscans}) which is given by the time it takes the strain pulse to propagate through Cu of thickness $d_\text{Cu}$ at the speed of sound $v_\text{s,Cu}=5.15\,\text{nm}/\text{ps}$~\cite{over1955}.
This strain pulse and the the strain pulse launched by the expansion of Ni are (partially) reflected at the surface and the substrate interface.
This results in periodically appearing signatures in the strain response such as at \qty{30}{ps} and $60\,\text{ps}$ for the $136\,\text{nm}$ thick Cu layer (see Fig.~\ref{fig:fig_2_delayscans}c) corresponding to the arrival of the Pt and the return of the Ni strain pulse, respectively.
For the thinnest Cu layers ($d_\text{Cu}\leq 47\,\text{nm}$), several periods of this periodic modulation are visible within the explored delay range.
The period is given by the acoustic thickness of the entire metallic heterostructure, i.e. the thickness of the layers divided by their sound velocity.
In addition to these coherent dynamics, the phonon temperatures of the metallic layers equilibrate on tens of picoseconds.
For the thickest Cu layer ($d_\text{Cu}=288\,\text{nm}$) this results in a slowly rising additional expansion of Ni since most of the energy initially remains within Pt and Cu after electron-phonon equilibration and is only transferred to the buried Ni layer via equilibrated electrons and phonons on tens of picoseconds.
In contrast, for the thin Cu layers ($d_\text{Cu}\leq47\,\text{nm}$) the expansion of Ni slightly decreases from $15$ to $85\,\text{ps}$ as a large fraction of the optically deposited energy is localized in Ni within the first picosecond.

We model the experimental strain employing the modular \textsc{Python} library \textsc{udkm1Dsim}~\cite{schi2021} with literature values for the thermophysical parameters~\cite{jare2024} (see supplementary material).
In the framework of our modeling, we first calculate the optical absorption by a transfer matrix algorithm and subsequently describe the spatio-temporal energy redistribution by a d2TM.
We explicitly calculate the energy redistribution via diffusion of electrons utilizing the temperature-dependent electronic heat conductivity $\kappa_\text{el}^0 \frac{T_\text{el}}{T_\text{ph}}$, which is required in order to account for the non-equilibrium among electrons and phonon directly after photo-excitation.
This modification of the equilibrium heat conductivity $\kappa_\text{el}^0$ by the electron ($T_\text{el}$) and phonon ($T_\text{ph}$) temperatures accounts for the temperature-dependent electronic heat capacity and scattering time, respectively~\cite{hohlfeld2000,pude2020}.
The resulting spatio-temporal energy density linearly translates to a stress on the lattice via Gr\"uneisen parameters~\cite{matt2023}.
It drives the spatio-temporal strain including the oscillatoric signatures of the launched strain pulses according to the linear one-dimensional, elastic wave equation.
Finally, dynamical x-ray scattering calculations considering the simulated strain yields layer-specific Bragg peaks whose shifts determine the layer-specific strain responses as in the experiment.
With a single set of parameters, our diffusive transport model matches the strain response of Ni (solid lines in Fig.~\ref{fig:fig_2_delayscans}) and Cu (solid lines in Fig.~S4) for low and high fluences and all four samples.
A small but noticeable deviation appears for $d_\text{Cu}=47\,\text{nm}$, where the experimental strain is systematically larger at early delays (see Fig.~\ref{fig:fig_2_delayscans}(b)). 
We attribute this deviation to a finite contribution of superdiffusive or ballistic electrons to the ultrafast energy transport, which is reasonable for a Cu thickness of about twice the inelastic mean free path ($\approx 25\,\text{nm}$)~\cite{schmuttenmaer1994}.
Because of a prominent direct excitation of Ni in the thinnest sample (see Fig.~\ref{fig:setup}(b)), this deviation is not observed there, indicating that ultrafast electronic energy transport only plays a minor role in that sample and we are not able to experimentally access its nature by our experiment.

\begin{figure}[!ht]
    \centering
    \includegraphics[width = \columnwidth]{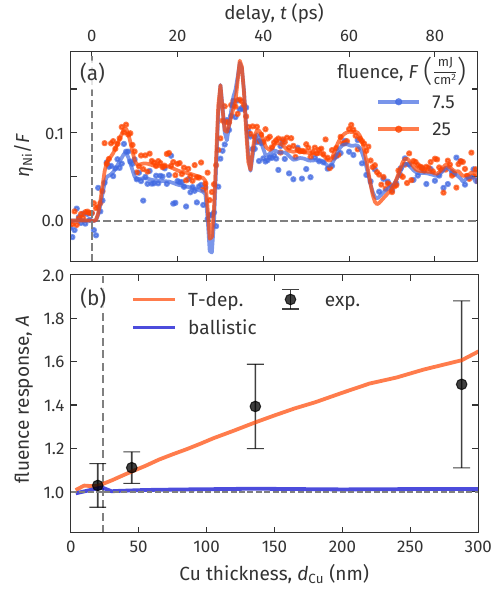}
    \caption{\label{fig:fig_3_nonlinearity}
    Nonlinear fluence dependence of the Ni strain in Pt-Cu-Ni heterostructures.
    (a) The transient Ni strain of the sample with \qty{136}{nm} Cu for the high and low fluence normalized to the respective incident fluence $F$.
    (b) The fluence response parameter $A$ calculated via Eq.~\eqref{eq:nonlinearity} as a function of Cu layer thickness for the experiment (symbols) and the diffusive (red) and quasi-ballistic (blue) transport models. 
    The measurements are best reproduced by the d2TM with temperature dependent electron conductivity.
    The errorbars are the standard error calculated from the strain values within the grey shaded areas in Fig.~\ref{fig:fig_2_delayscans}.
    The dashed vertical line indicates the inelastic mean free path of electrons in Cu.} 
\end{figure}

In the following, we discuss the fluence dependence of the ultrafast energy transport through Cu via hot electrons.
Figure~\ref{fig:fig_3_nonlinearity}(a) exemplarily displays the transient Ni strain normalized to the corresponding excitation fluence for the sample with \qty{136}{nm} of Cu for the high (red symbols and line) and low (blue symbols and line) fluence.
While the fluence-normalized strain for both measurements matches at $t>80\,\text{ps}$ (i.e. linear fluence dependence), we observe a larger strain for the high fluence at $t<26\,\text{ps}$.
The initial Ni expansion $\eta_\text{Ni}(t_1)$ averaged from the delay of established expansion at the speed of sound ($2.7\,\text{ps}$) until the entrance of the Pt strain pulse at $d_\text{Cu}/v_\text{s,Cu}$ (first gray box in Fig.~\ref{fig:fig_2_delayscans}) is linearly related to the energy density deposited to Ni within the first hundreds of femtoseconds via electronic energy transport from Pt through Cu or by direct optical absorption.
On the other hand, $\eta_\text{Ni}(t_2)$ the expansion averaged from $84$ to $89\,\text{ps}$ (second gray box in Fig.~\ref{fig:fig_2_delayscans}) is given by the total optically deposited energy that is spread across the entire metallic heterostructure which depends linearly on the fluence.
Therefore, the ratio $R=\eta_\text{Ni}(t_1)/\eta_\text{Ni}(t_2)$ is a measure of how efficient energy is rapidly transported through the Cu layer via hot electrons.
Note, the delay range $t_2$ is chosen to minimize the strain contributions from the coherently driven propagating strain pulses and the small thermal conductivity of the glass substrate ensures that only a negligible fraction of energy is transported to the substrate on the recorded timescales.

The non-linearity of the rapid Ni expansion ($\eta_\text{Ni}(t_1)$) shown in Fig.~\ref{fig:fig_3_nonlinearity}(a) translates to different ratios $R^\text{high}$ and $R^\text{low}$ for high and low fluences, respectively.
Accordingly, we define a \emph{fluence response parameter}, $A$, by
\begin{align}
    A = \frac{R^\text{high}}{R^\text{low}} = \frac{\eta_\text{Ni}^\text{high}(t_1)}{\eta_\text{Ni}^\text{low}(t_1)} \cdot \frac{\eta_\text{Ni}^\text{low}(t_2)}{\eta_\text{Ni}^\text{high}(t_2)}
    \label{eq:nonlinearity}
\end{align}
with the Ni strain for the high fluence $\eta_\text{Ni}^\text{high}$ and low fluence $\eta_\text{Ni}^\text{low}$ averaged over the two distinct delay ranges, respectively.
Our experimental $A$ (symbols in Fig.~\ref{fig:fig_3_nonlinearity}(b) significantly increases with increasing $d_\text{Cu}$ for $d_\text{Cu} \geq 47\,\text{nm}$ indicating a fluence-dependent efficiency of the ultrafast electronic energy transport.
This is in good agreement with the diffusive electron transport in the framework of the d2TM (red solid line) that becomes non-linear by the temperature-dependent electronic thermal conductivity $\kappa_\text{el}^0 \frac{T_\text{el}}{T_\text{ph}}$ accounting for the laser-induced electron-phonon nonequilibrium.
In contrast, a quasi-ballistic transport model (blue solid line) depends linearly on the fluence ($A_\text{bal}=1$) due to an independent propagation speed of the heat carrying electrons.
We approximates the ballistic electron transport within our diffusive modeling framework by an instantaneous equilibration of the electron temperature across the metallic heterostructure~\cite{pude2020,ashok2025} to account for the propagation of the heat carrying electrons at Fermi velocity.
In case of UXRD experiments, this is a useful approximation as the short timescale of the ballistic transport via electrons propagating at Fermi velocity is much faster than establishing the lattice expansion at the speed of sound~\cite{matt2023}.
We emphasize that ballistic transport not only results in a deviating $A$ from the experimental values beyond the error bars for $d_\text{Cu} \leq 47\,\text{nm}$ but also significantly overestimates the initial expansion of Ni (see Fig.~S6).
Thus, the characteristic nonlinear fluence-dependence of the initial expansion of the buried Ni layer $\eta_\text{Ni}(t_1)$ serves as an unambiguous experimental marker of diffusive energy transport via electrons that dominates for Cu thicknesses above $47\,\text{nm}$.

In conclusion, we identified the dominant contribution of diffusive electronic energy transport through Cu layers of $d_\text{Cu} \geq 47\,\text{nm}$ by tracking the fluence-dependent strain response of a buried Ni detection layer within a Pt-Cu-Ni heterostructure via UXRD.
We demonstrated that the fluence-dependent efficiency of the electronic energy transport, i.e. the fluence response parameter $A$, unambiguously discriminates diffusive and ballistic energy transport.
It can be experimentally determined from the fluence-dependent ratio of the initial rapid expansion of a buried detection layer and its expansion after equilibration of the phonon temperatures across the metallic heterostructure.
This approach is not based on the different transport velocities of diffusive and ballistic transport and therefore does not require a quasi-instantaneous response of the experimental observable to the deposited energy which makes it broadly applicable.
The linear relation between the lattice expansion and the energy stored within a layer renders UXRD an ideal technique to characterize the fluence-dependence of the transport directly from the experimental response.
We expect ballistic transport to dominate the ultrafast energy transport in case of propagation layers thinner than the inelastic mean free path of electrons.
In our experiment this regime was not accessible due to a dominant direct photo-excitation of the Ni detection layer.
In future experiments, tailoring the absorption by selecting an excitation in the UV or XUV range with significantly shorter penetration depth would make the nature of the energy transport through thinner propagation layers experimentally accessible.
Our results, which we expect to be transferable to other noble metals such as Au or Ag, represent important insight for future applications that rely on the transport of electrons well above the Fermi energy through into functional layers embedded in metallic heterostructures.

\begin{acknowledgments}
J.J., M.M. and D.S. would like to thank the Leibniz Association for funding through the Leibniz Junior Research Group J134/2022.
L.M. acknowledges funding by the Deutsche Forschungsgemeinschaft (DFG, German Research Foundation) – CRC/SFB 1636 – Project ID 510943930 - Project No. A01.
C.W. acknowledges the DFG for financial support Project No. 328545488—TRR 227, project A10.
M.H. acknowledges financial support from the France 2030 government grants PEPR SPIN (ANR-22-EXSP 0007).
This work was supported by the ANR through the France 2030 government grants TOAST (ANR-22-EXSP 0003) and PEPR SPIN – SPINMAT ANR-22-EXSP- 0007. 
J-E.P. would like to thank the European X-ray Free-Electron Laser Facility for funding. J.L. acknowledges the support from the Swedish Research Council (VR, Grant No. 2023-05136), and Olle Engkvists Stiftelse (Grant 238-0012).
We acknowledge the MAX IV Laboratory for beamtime on the FemtoMAX beamline under proposal 20240531. Research conducted at MAX IV, a Swedish national user facility, is supported by Vetenskapsrådet (Swedish Research Council, VR) under contract 2018-07152, Vinnova (Swedish Governmental Agency for Innovation Systems) under contract 2018-04969 and Formas under contract 2019-02496.
\end{acknowledgments}

\bibliography{literature.bib}

\end{document}


\preprint{APS/123-QED}

\title{Supplementary material to: Experimental evidence of dominant ultrafast diffusive energy transport by hot electrons in Cu}

\author{J.~Jarecki}
\affiliation{Max-Born-Institut f\"ur Nichtlineare Optik und Kurzzeitspektroskopie, 12489 Berlin, Germany}
\email{jarecki@mbi-berlin.de}

\author{L.~Mehner}
\affiliation{Institut f\"ur Physik und Astronomie, Universit\"at Potsdam, 14476 Potsdam, Germany}

\author{M.~Mattern}
\affiliation{Max-Born-Institut f\"ur Nichtlineare Optik und Kurzzeitspektroskopie, 12489 Berlin, Germany}
\email{mattern@mbi-berlin.de}

\author{A.~Jurgilaitis}
\affiliation{MAX IV Laboratory, Lund University, PO Box 118, 221 00 Lund, Sweden}

\author{S.-P.~Zeuschner}
\affiliation{Institut f\"ur Physik und Astronomie, Universit\"at Potsdam, 14476 Potsdam, Germany}

\author{B.~Ahn}
\affiliation{MAX IV Laboratory, Lund University, PO Box 118, 221 00 Lund, Sweden}

\author{F.~Baltrusch}
\affiliation{Institut f\"ur Physik und Astronomie, Universit\"at Potsdam, 14476 Potsdam, Germany}

\author{J. C.~Ekstr\"om}
\affiliation{MAX IV Laboratory, Lund University, PO Box 118, 221 00 Lund, Sweden}

\author{D.~Kroon}
\affiliation{MAX IV Laboratory, Lund University, PO Box 118, 221 00 Lund, Sweden}

\author{M.~Herzog}
\affiliation{Institut f\"ur Physik und Astronomie, Universit\"at Potsdam, 14476 Potsdam, Germany}

\author{C.~Walz}
\affiliation{Institut f\"ur Physik und Astronomie, Universit\"at Potsdam, 14476 Potsdam, Germany}

\author{F.-C.~Weber}
\affiliation{Institut f\"ur Physik und Astronomie, Universit\"at Potsdam, 14476 Potsdam, Germany}
\affiliation{Helmholtz-Zentrum Berlin f\"ur Materialien und Energie GmbH, Wilhelm-Conrad-R\"ontgen Campus, BESSY II, 12489 Berlin, Germany}

\author{J.~Larsson}
\affiliation{Department of Physics, Lund University, PO Box 118, Lund 221 00, Sweden}
\affiliation{MAX IV Laboratory, Lund University, PO Box 118, 221 00 Lund, Sweden}
\affiliation{ LINXS Institute of advanced Neutron and X-ray Science, Mesongatan 4, 224 84 Lund, Sweden}

\author{M.~Hehn}
\affiliation{Institut Jean Lamour (UMR CNRS 7198), Universit\'e Lorraine,  54000 Nancy,   France}

\author{J.-E.~Pudell}
\affiliation{European X-ray Free-Electron Laser Facility, 22869 Schenefeld, Germany}

\author{D.~Schick}
\affiliation{Max-Born-Institut f\"ur Nichtlineare Optik und Kurzzeitspektroskopie, 12489 Berlin, Germany}

\author{A.~von Reppert}
\affiliation{Institut f\"ur Physik und Astronomie, Universit\"at Potsdam, 14476 Potsdam, Germany}

\author{M.~Bargheer}
\affiliation{Institut f\"ur Physik und Astronomie, Universit\"at Potsdam, 14476 Potsdam, Germany}
\affiliation{Helmholtz-Zentrum Berlin f\"ur Materialien und Energie GmbH, Wilhelm-Conrad-R\"ontgen Campus, BESSY II, 12489 Berlin, Germany}
\date{\today} 
\date{\today} 
\maketitle
\renewcommand{\thefigure}{S\arabic{figure}} 
\renewcommand{\thesection}{S\arabic{section}} 

\section{Ultrafast x-ray diffraction experiments}

The ultrafast x-ray diffraction (UXRD) experiments were mainly conducted at the FemtoMAX beamline at the MAX IV laboratory~\cite{enqu2018}, and complemented by additional UXRD measurements at a table-top laser-driven plasma x-ray source (PXS)~\cite{zamp2009,schi2012}.
At both setups, we study the laser-induced lattice response of the metallic heterostructure employing femtosecond x-ray probe pulses with $8\,\text{keV}$ photon energy, $<100\ \text{fs}$ (FWHM) pump pulse with a central wavelength of $800\,\text{nm}$ and comparable pixelated position-sensitive \textsc{Pilatus} detectors via reciprocal space slicing~\cite{zeus2021}.
While at FemtoMAX the sample is excited by near-infrared $50 \,\text{fs}$ pump pulses at a repetition rate of $10\,\text{Hz}$ incident at around $24^\circ$, at the PXS the sample is excited by $60\,\text{fs}$ pump pulses with a repetition rate of $1\,\text{kHz}$ incident under $42^\circ$.
These slight differences in pulse durations do not influence our results as we focus on dynamics on picosecond timescales and both are well below the electron-phonon coupling times in the involved metals.
The different angle of incidence does not significantly change the absorption profile according to our multilayer model.
In order to characterize the nonlinearity of the energy transport, we measured each sample at a low and high excitation fluence $F$.
Details about the experimental parameters are given in Tab.~\ref{tab:experiment}.
\begin{table}[H]
\centering
\renewcommand{\arraystretch}{1.3}
\setlength{\tabcolsep}{10pt}
\begin{tabular}{l|r|c|c}
    setup & sample & low fluence & high fluence \\
    \hline
    FemtoMAX & \qty{47}{nm} Cu & \qty{7.5}{mJ\per cm\squared} & \qty{25.0}{mJ\per cm\squared} \\
             & \qty{136}{nm} Cu & - & \qty{25.0}{mJ\per cm\squared} \\ 
             & \qty{288}{nm} Cu & \qty{7.5}{mJ\per cm\squared} & \qty{25.0}{mJ\per cm\squared} \\
    \hline
    PXS      & \qty{22}{nm} Cu & \qty{3.0}{mJ\per cm\squared} & \qty{10.0}{mJ\per cm\squared} \\
             & \qty{136}{nm} Cu & \qty{7.5}{mJ\per cm\squared} & - \\
\end{tabular}
\label{tab:experiment}
\caption{Summary of experiments and corresponding excitation fluences.}
\end{table}

\begin{figure}[H]
    \centering
    \includegraphics{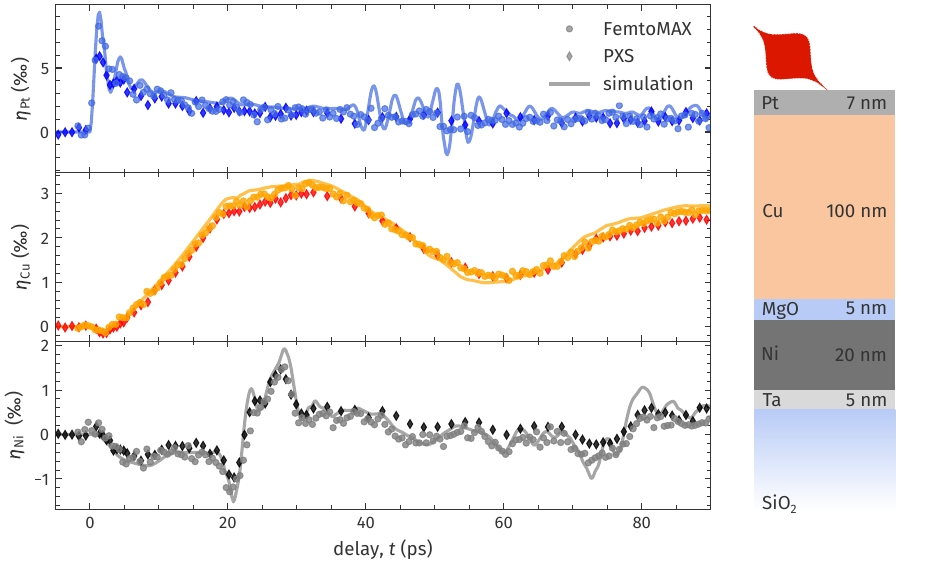}
    \caption{
    Calibration measurement to showcase the comparability of the FemtoMAX and PXS UXRD experiments. 
    The comparison of the strain at FemtoMAX (symbols) and the PXS (dashed line) for Pt, Cu and Ni from top to bottom demonstrates the comparability of both experiments as they provide similar results compared to the strain simulation (solid line).}
    \label{fig:S1_calib}
\end{figure}

To confirm that the minor differences in the setups do not influence the experimental results, we performed calibration measurements on a reference sample built of Pt (\qty{7}{nm}), Cu (\qty{100}{nm}), MgO (\qty{5}{nm}) and Ni (\qty{20}{nm}) grown on a \qty{5}{nm} Ta buffer layer on a glass substrate. 
The time-dependent strain response of the main three metal layers is shown in Fig.~\ref{fig:S1_calib}. 
The experimental strain measured at FemtoMAX (symbols) is in good agreement with the strain measured at the PXS (dashed lines) for all layers. 
We also plot the corresponding simulation of the strain (solid lines), which supports the reliability of the calibration. These results confirm the comparability of the two experiments and that no significant differences are introduced.

\newpage

\section{Static Sample characterization}
\begin{figure}[H]
    \centering
    \includegraphics[width=1\textwidth]{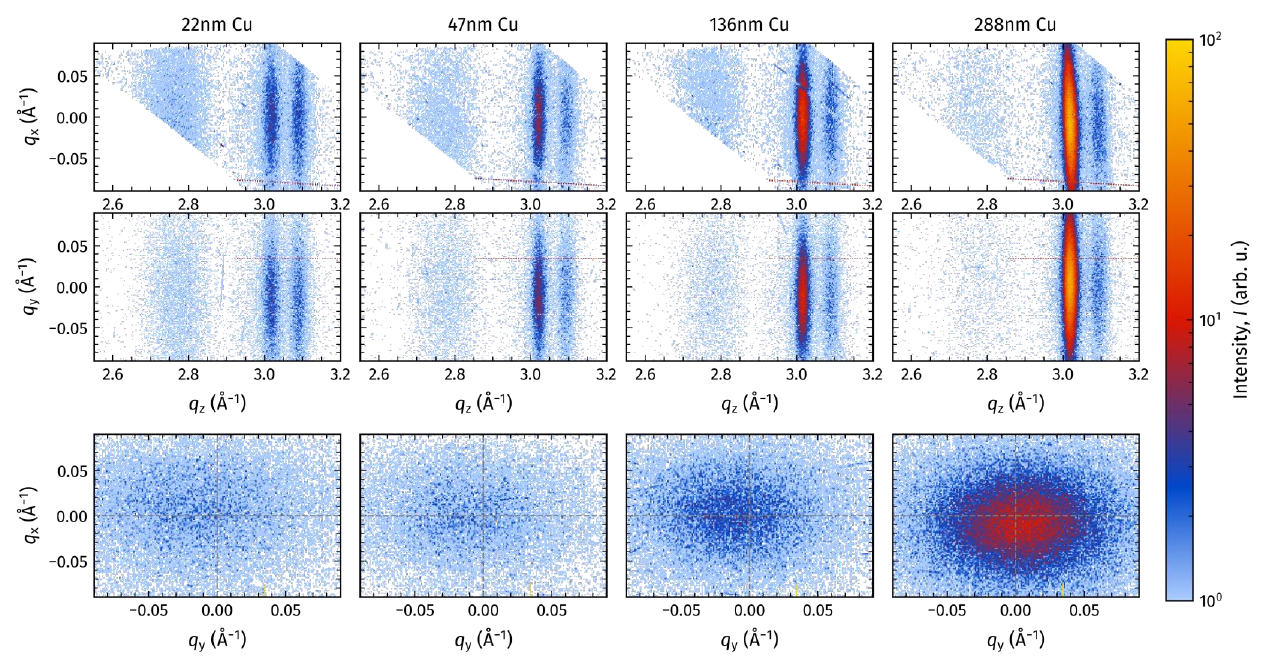}
    \caption{Reciprocal space maps (RSM) of the samples with \qty{22}{nm}, \qty{47}{nm}, \qty{136}{nm}, and \qty{288}{nm} Cu from left to right.
    The RSMs were recorded by rotating the sample and detector in symmetric $\theta-2\theta$-scans and display the diffracted intensity as a function of the out-of-plane and in-plane reciprocal coordinates $q_z$ and $q_x$, $q_y$, respectively.
    We observe clear signatures of the Cu and Ni (111) Bragg peaks around $3\,\text{\AA}^{-1}$ and $3.1\,\text{\AA}^{-1}$, respectively.
    The intensity of the Cu Bragg peak increases with increasing Cu layer thickness.
    In contrast, the intensity of the Ni Bragg peak barely differs among the different samples indicating the comparability of the sample series.
    The intensity of the very broad Bragg peak of the \qty{5}{nm} Pt capping layer is rather low and even decreases with increasing Cu thickness which makes it inaccessible in the UXRD experiment.
    The large widths of all Bragg peaks along the in-plane reciprocal coordinates $q_x$ and $q_y$ indicates the textured mosaic growth of the layers.}
    \label{fig:S2_rsms}
\end{figure}

\begin{figure}[H]
    \centering
    \includegraphics[width=1\textwidth]{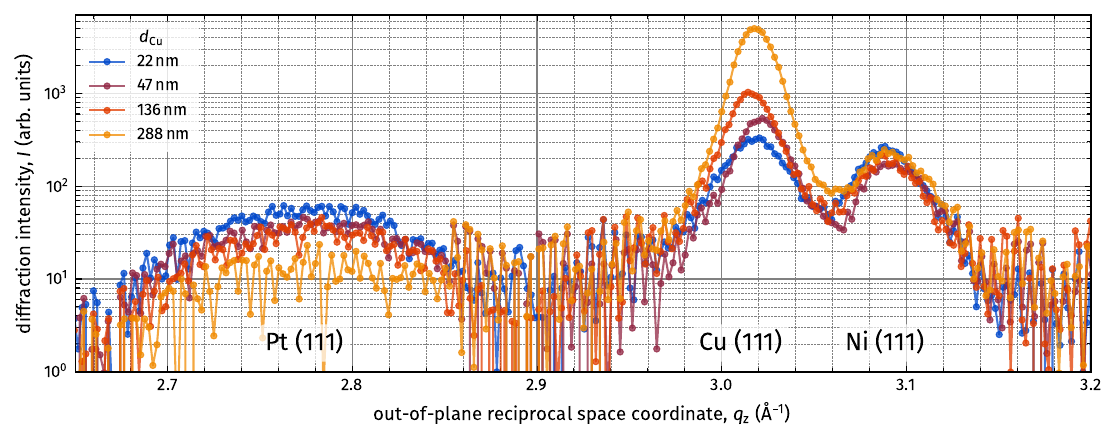}
    \caption{Integrated diffracted intensity along the out-of-plane reciprocal coordinate $q_z$ for all four samples.
    The most prominent change among the samples is the intensity change of the Cu Bragg peak.
    In contrast the Ni Bragg peaks are nearly identical for all samples confirming that the samples are comparable in our experiment.
    For a Cu thickness of \qty{136}{nm}, and \qty{288}{nm} the Pt Bragg peak barely exceeds the noise background.}
    \label{fig:S3_Iqz}
\end{figure}

\newpage

\section{Strain dynamics of the Cu transport layer}

\begin{figure}[H]
    \centering
    \includegraphics[width=0.5\textwidth]{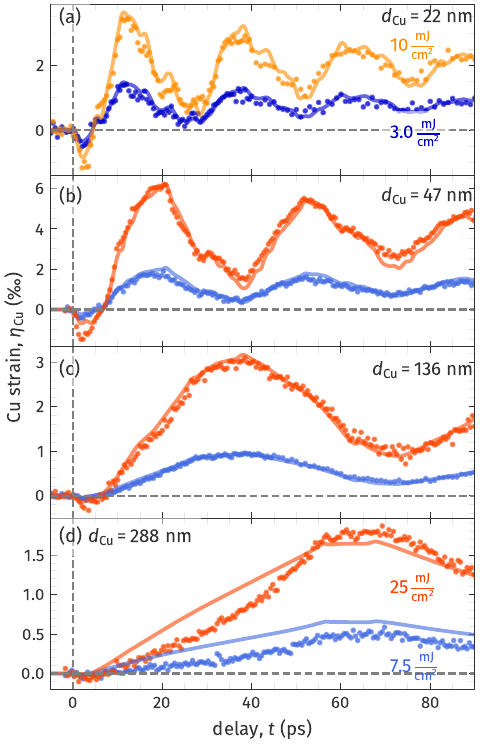}
    \caption{Laser-induced strain response of the Cu transport layer.
    The transient strain of the Cu layer (symbols) for \qty{22}{nm} (a), \qty{47}{nm} (b), \qty{136}{nm} (c) and \qty{288}{nm} (d) thickness upon excitation with low (blue) and high (red) fluence.
    Irrespective of the Cu thickness, we observe an initial compression for the high fluence that originates from the expansion of the neighboring Pt and Ni layers.
    As discussed in the main text, the diffusive energy transport through Cu into Ni non-linearly increases with fluence which results in the more pronounced compression of Cu.
    Subsequently, Cu expands due to energy transport of equilibrated electrons and phonons and the strain shows an oscillation whose period is given by the acoustic thickness of the metallic heterostructure (thickness of each material weighted with its sound velocity) and accordingly increases with increasing Cu thickness $d_\text{Cu}$.
    Our diffusive two-temperature model (solid lines) of the strain response of the metallic heterostructure, that matches the experimental results in Ni (see Fig. 2), also matches the response of Cu with the same set of parameters.}
    \label{fig:S4_cu_strain}
\end{figure}

\newpage

\section{Thermo-elastic parameters of the Pt-Cu-Ni heterostructure}

The thermo-physical parameters used in the modeling are given the table below (Tab.~\ref{tab:tab_1_sim_param}). 
We use a single set of parameters to describe all samples with the different Cu layer thicknesses. 
The exact same literature parameters have been further used before to describe the temperature and strain dynamics of similar Pt-Cu-Ni heterostructures with slightly different sample designs~\cite{jare2024}.
\begin{table*}[h]
\centering
\begin{ruledtabular}
\begin{tabular}{l c c c c c c}
 & Pt & Cu & Ni & Ta & glass \\
 \hline
$\Tilde{n}$ & 0.576+8.078i~\cite{wern2009} & 0.105+5.141i~\cite{wern2009} & 2.322+8.882i~\cite{wern2009} & 0.992+7.293i~\cite{wern2009} & 1.5~\cite{corn2002}\\
$\gamma\textsuperscript{S}$ (mJ\,cm\textsuperscript{-3}\,K\textsuperscript{-2})& 0.73~\cite{hohlfeld2000} & 0.10~\cite{lin2008} & 1.06~\cite{hohlfeld2000} & 0.38~\cite{bodr2013}  & -\\
$C\textsubscript{ph}$ (J\,cm\textsuperscript{-3}\,K\textsuperscript{-1}) & 2.85~\cite{shay2016} & 3.44 & 3.94~\cite{mesc1981}  & 2.33~\cite{bodr2013} & 1.80~\cite{corn2002}\\
$\kappa_\text{el}^0$ (W\,m\textsuperscript{-1}\,K\textsuperscript{-1}) & 66~\cite{dugg1970} & 396~\cite{hohlfeld2000} &  81.4~\cite{hohlfeld2000} & 52.0  & -\\
$\kappa\textsubscript{ph}$ (W\,m\textsuperscript{-1}\,K\textsuperscript{-1}) &5.0~\cite{dugg1970} & 5.0&  9.6~\cite{hohlfeld2000} & 5.0 & 1.0~\cite{corn2002}\\
$g$ (PW\,m\textsuperscript{-3}\,K\textsuperscript{-1})& 375~\cite{zahn2021} & 95~\cite{lin2008} & 360~\cite{lin2008} & 100  &-\\
$\rho$ (g\,cm\textsuperscript{-3}) & 21.45 & 8.96 & 8.91 & 16.68 & 2.54~\cite{corn2002}\\
$v\textsubscript{S}$ (nm\,ps\textsuperscript{-1})& 4.2~\cite{farl1966} & 5.2~\cite{over1955} & 6.3~\cite{neig1952} & 4.2~\cite{feat1963} & 5.7~\cite{corn2002} \\
$\Gamma$\textsubscript{el} & 2.4~\cite{kris1979}(1.2) & 0.9~\cite{kris1979} & 1.4~\cite{wang2008} & 1.3~\cite{kris1979} & - \\
$\Gamma$\textsubscript{ph} & 2.6~\cite{nix1942} & 2.0~\cite{nix1941} & 1.8~\cite{wang2008} & 1.6~\cite{kris1979} & 0.3~\cite{corn2002}\\
\end{tabular}
\end{ruledtabular}
\caption{Literature values for the physical parameters of the strain model. 
The complex refractive index $\Tilde{n}$, the Sommerfeld constant $\gamma\textsuperscript{S}$, the specific heat of the phonons $C\textsubscript{ph}$, the electron-phonon coupling constant $g$, and the electron $\kappa_e^0$ and phonon $\kappa\textsubscript{ph}$ heat conductivity determine the spatio-temporal energy distribution upon laser-excitation. 
The subsystem-specific Gr\"uneisen parameters $\Gamma$\textsubscript{el} and $\Gamma$\textsubscript{ph} linearly relate the spatio-temporal energy density to an elastic stress on the lattice driving a quasi-static expansion and strain pulses propagating with sound velocity $v\textsubscript{S}$ according to the elastic wave equation.
Values in brackets are optimized values for the strain modeling.}
\label{tab:tab_1_sim_param}
\end{table*}

\section{Direct optical excitation of the buried Ni detection layer}
\begin{figure}[h]
    \centering
    \includegraphics[width=1\textwidth]{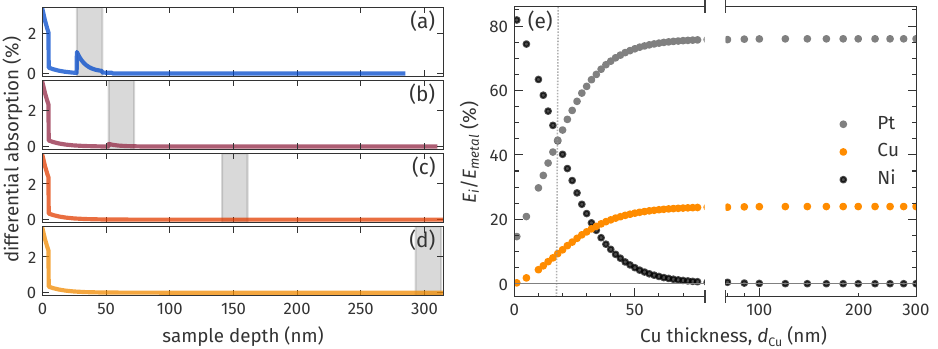}
    \caption{Optical excitation of the Pt-Cu-Ni heterostructures. 
    The calculated absorption profile of the \qty{800}{nm} pump pulses in the framework of a multilayer transfer matrix model for the samples with a \qty{22}{nm} (a), \qty{47}{nm} (b), \qty{136}{nm} (c) , and \qty{288}{nm} (d) Cu layer for a \qty{45}{\degree} angle of incidence. 
    The gray shaded areas indicate the Ni layer.
    While the Ni layer is directly excited by the \qty{800}{nm} pump laser quite significantly for a Cu thickness of $d_\text{Cu} = 22\,$nm, direct absorption by Ni can be neglected for larger Cu thicknesses.
    (e) The energy deposited by the femtosecond laser pulses to the Pt, Cu and Ni layer normalized to the totally deposited energy $E_\text{metal}$ as a function of Cu layer thickness.
    The vertical dashed line denotes the Cu thickness where the absorption in Pt surpasses that of Ni and therefore transport processes start to play a role for the transient energy stored in Ni.
    For $d_\text{Cu}>70\,\text{nm}$, the direct optical excitation in Ni becomes negligible and energy transport via electrons governs the lattice expansion of the buried Ni detection layer.
    }
    \label{fig:S5_absorption}

\end{figure}

\newpage

\section{Comparison of the diffusive and ballistic model}

\begin{figure}[H]
    \centering
    \includegraphics[width=1\textwidth]{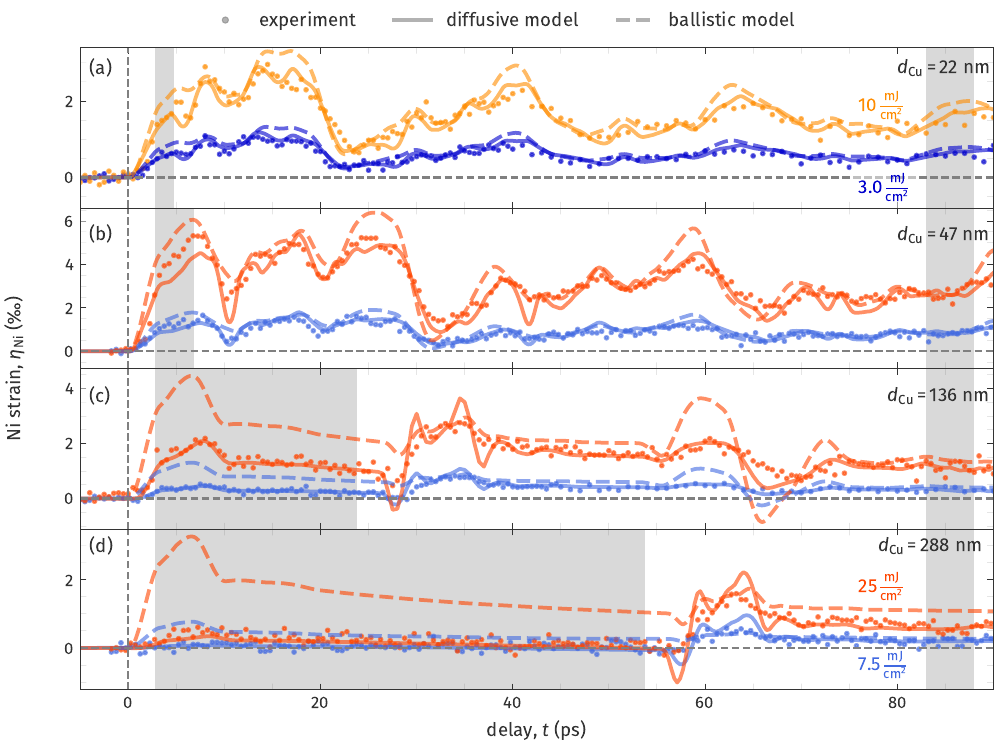}
    \caption{Modeled transient Ni strain in case of diffusive and ballistic energy transport via hot electrons.
    The transient strain of the Ni layer (symbols) for \qty{22}{nm} (a), \qty{47}{nm} (b), \qty{136}{nm} (c) , and \qty{288}{nm} (d) Cu thickness upon excitation with low (blue) and high (red) fluence in comparison to the diffusive transport model (solid lines) and the quasi-ballistic model (dashed lines).
    The gray shaded areas denote the delay ranges used to calculate the fluence response parameter $A$ in the main manuscript.
    As highlighted in the main manuscript the diffusive model matches the experimental data very well, the ballistic model, however, significantly overestimates the Ni strain at early delays for large Cu thicknesses in particular.
    While for \qty{22}{nm} of Cu the differences between the simulated strain response for both transport models are small indicating the negligible role of transport and the dominating effect of direct optical absorption in Ni, the increasing overestimation of the ballistic model confirms the minor contribution of ballistic electrons to the energy transfer for Cu thicknesses substantially above the inelastic mean free path of Cu.
    A finite effect of ballistic electrons is suspected for \qty{47}{nm} of Cu, where the experimental strain response exceeds the diffusive model but does not reach the expected strain from the ballistic model. 
    }
    \label{fig:models}

\end{figure}
A description of the general workflow of our modeling as well as details about the simulation of the diffusive transport model are given in the main manuscript.
The modeling of the quasi-ballistic scenario is done within the same framework employing the modular \textsc{Python} library \textsc{udkm1Dsim}~\cite{schi2021} with literature values for the thermophysical parameters (see Tab.~\ref{tab:tab_1_sim_param}).
However, we approximate the rapid energy redistribution associated by ballistic electrons propagating at Fermi velocity by an instantaneous equilibration of the electron temperature across the metallic heterostructure.
For that, we first calculate the total energy $E_\text{tot}$ which is deposited by the optical pump pulse using a transfer matrix algorithm and subsequently the corresponding electron temperature $T_0^\text{ex}$
\begin{equation}
    T_0^\text{ex} = \sqrt{2\,\frac{\rho_\text{tot}}{\gamma^\text{tot}} + T_0^2}
\end{equation}
using the total energy density $\rho_\text{tot}$, the initial electron temperature of the electrons in equilibrium $T_0$ and an effective electronic heat capacity for the metal stack
\begin{equation}
    \gamma^\text{tot} = (\gamma^\text{Pt} \cdot d_\text{Pt} + \gamma^\text{Cu} \cdot d_\text{Cu} + \gamma^\text{Ni} \cdot d_\text{Ni} + \gamma^\text{Ta} \cdot d_\text{Ta}) / d_\text{tot}
\end{equation}
with $\gamma^i$ and $d^i$ the Sommerfeld constant and the thickness of the constituent metal layers, respectively.
This means, we set $T_0^\text{ex}$ as spatially homogeneous initial electron temperature after the laser pulse arrival and describe the subsequent spatio-temporal energy redistribution again by the same diffusive two-temperature model (d2TM) as described in the main manuscript.

\bibliography{literature.bib}